\documentclass[preprint,showpacs,preprintnumbers,amsmath,amssymb]{revtex4}
\usepackage{graphicx, epsfig, dcolumn}
\begin{document}

%\preprint{ }

\title{Local Atomic Structure of Martensitic Ni$_{2+x}$Mn$_{1-x}$Ga: An EXAFS Study}

\author{P. A. Bhobe} 
\author{K. R. Priolkar}\email[corresponding author: ]{krp@unigoa.ac.in} \author{P. R. Sarode}
\affiliation{Department of Physics, Goa University, Goa, 403 206 India.}

\date{\today}

\begin{abstract}
The local atomic structure of Ni$_{2+x}$Mn$_{1-x}$Ga with 0 $\le$ $x$ $\le$ 0.16 alloys was explored using Mn and Ga K-edge Extended X-ray Absorption Fine Structure (EXAFS) measurement. Inorder to study the atomic re-arrangements that occur upon martensitic transformation, room temperature and low temperature EXAFS were recorded. The changes occurring in the L2$_1$ unit cell and the bond lengths obtained from the analysis enables us to determine the modulation amplitudes over which the constituent atoms move giving rise to shuffling of the atomic planes in the modulated structure. The EXAFS analysis also suggests the changes in hybridization of Ga-$p$ and Ni-$d$ orbitals associated with the local symmetry breaking upon undergoing martensitic transition.     
\end{abstract}
\maketitle

\section{\label{sec:level1}Introduction}
Martnesitic transformations are first-order, displacive, solid-solid phase transformation taking place upon cooling below a characteristic temperature T$_M$ from a high symmetry initial phase to a low-symmetry structure.  Ni$_2$MnGa exhibits martensitic transformation upon cooling through 220K \cite{Web}. Moreover, it is ferromagnetic with a Curie temperature, T$_C \sim$ 370K making it a technologically important magnetic shape memory alloy. Alot of theoretical\cite{ayu1, god, ayu2, mac, lee, bun, zay1, zay-role, zay-prb-72, bar} and experimental studies \cite{mart, ooi, wirth, sov, vasil1, pab, kuo} that reflect the Fermi surface character of Ni$_2$MnGa upon undergoing martensitic transformation have been carried out. The early neutron diffraction study \cite{Web} determined the structure of the stoichiometric Ni$_2$MnGa to be cubic L2$_1$ Heusler type with a = 5.825\AA~ and a complete crystallographic structure determination of different martensitic phases has also been reported\cite{brow14}. Many other investigations have been carried out on the near-stoichiometric alloys confirming that the martensitic structure is a body centered tetragonal distortion of the initial cubic lattice. The low temperature crystal structure of the non-stoichiometric Ni-Mn-Ga alloys revealed that there exists different intermartensitic transformations as the lattice is subjected to periodic shuffling of the (110) planes along the [1$\bar 1$0]$_P$ direction of the initial cubic system \cite{mart} with modulation period dependent on the composition as summarized in \cite{pons}.  Inelastic scattering measurements made on the single crystals showed the presence of precursor effect above T$_M$ that gives rise to softening of [$\zeta$ $\zeta$ 0]TA$_2$ phonon mode at wave vector $\zeta_0$$\sim$ 0.33 of the reciprocal lattice \cite{Zheu, stur, com}. A large softening of certain elastic constants $C'$ = $\frac{1}{2}$($C_{11} - C_{12}$) takes place at this intermediate transition\cite{mano}. The dependence of $C'$ on applied magnetic field proves the magnetoelastic origins of these interactions \cite{plan}. Extensive studies in the past have attributed the structural transformations to the phonon anomalies occurring in the parent phase \cite{shap1, shap2}. Here, an incomplete softening of the [$\zeta$ $\zeta$ 0]TA$_2$ phonon mode at a particular wave vector $\zeta_0$(corresponding to the periodicity of the martensitic phase) with displacement along the [110] direction takes place. Such a phonon softening is believed to be due to contribution from electron-lattice coupling and nesting of the Fermi surface \cite{zhao1, zhao2, dug}. In spite of intense efforts,     the underlying mechanism giving rise to such a phase transformation is still not well understood. The nature of modulations forming the super structures and the driving force for the martensitic transformation in these alloys is currently at debate. Recent calculations by \cite{zay1, zay-role} indicate the importance of modulated structure and the shuffling of atomic planes in stabilizing the martensitic structure in stoichiometric and non-stoichiometric alloys. The stability of the structure is associated with a dip in the minority-spin density of states (DOS) at the Fermi level, related to the formation of hybrid states of Ni d and Ga p minority-spin orbitals \cite{zay-prb-72, bar}. The shuffles in these alloys are believed to be due to two different effects - modulations and tetrahedral distortions \cite{zay-role}. Also it is predicted that the amplitude of modulations are different for Mn-Ga and Ni planes \cite{zay1}. Therefore a precise knowledge of the changes occurring in the local structure of constituent atoms is fundamental in understanding the mechanism involved in martensitic transformation. EXAFS is an ideal tool to study such transformations by making a comparative study of the local structure in austinitic and martensitic phases. It is with this objective that the present investigation was undertaken.

In the present work, we report our study on Mn and Ga K-edge XAFS carried out at room temperature and liquid Nitrogen temperature in the Ni-Mn-Ga system to explore the changes in local environment around these metal ions in the austenitic and martensitic phase. We have carried out the measurements on the alloy compositions: Ni$_{2+x}$Mn$_{1-x}$Ga with $x$ = 0, 0.1, 0.13, and 0.16. The T$_M$ in this series is known to increase systematically from 220K to 315K with increasing Ni content\cite{vasil2}. Essentially, samples with $x$ = 0, and 0.10 are in the austinitic phase at room temperature, whereas $x$ = 0.13 undergoes a transition at $\sim$295K and $x$ = 0.16 represents the martensitic phase with T$_M$ = 315K. At liquid Nitrogen temperature all the four samples are in the martensitic phase.

\section{\label{sec:level1}Experimental Details}
Polycrystalline ingots of Ni$_{2+x}$Mn$_{1-x}$Ga 0 $\le$ $x$ $\le$ 0.16 were prepared by conventional arc-melting method in Argon atmosphere. The ingots were sealed in an evacuated quartz ampoule and annealed at 1000K for 48 hours followed by quenching in cold water to improve the homogeneity of the samples. A piece of the ingot was crushed into fine powder and further annealed at 1000K for 24 hours in an evacuated quartz tube inorder to remove any internal stress. Room temperature powder X-ray diffraction patterns recorded on Rigaku D-MAX IIC diffractometer with Cu K$\alpha$ radiation indicated the samples to be phase pure with L2$_1$ structure for $x$ = 0, 0.1 and a modulated tetragonal structure for $x$ = 0.13 and 0.16 samples. Energy dispersive X-ray (EDX) analysis confirmed the compositions to be nominal. The temperature dependent magnetic susceptibility and four probe resistivity measurements yielded $T_M$ and $T_C$ values same as those obtained by \cite{vasil2}.

\begin{figure}[h]
\epsfig{file=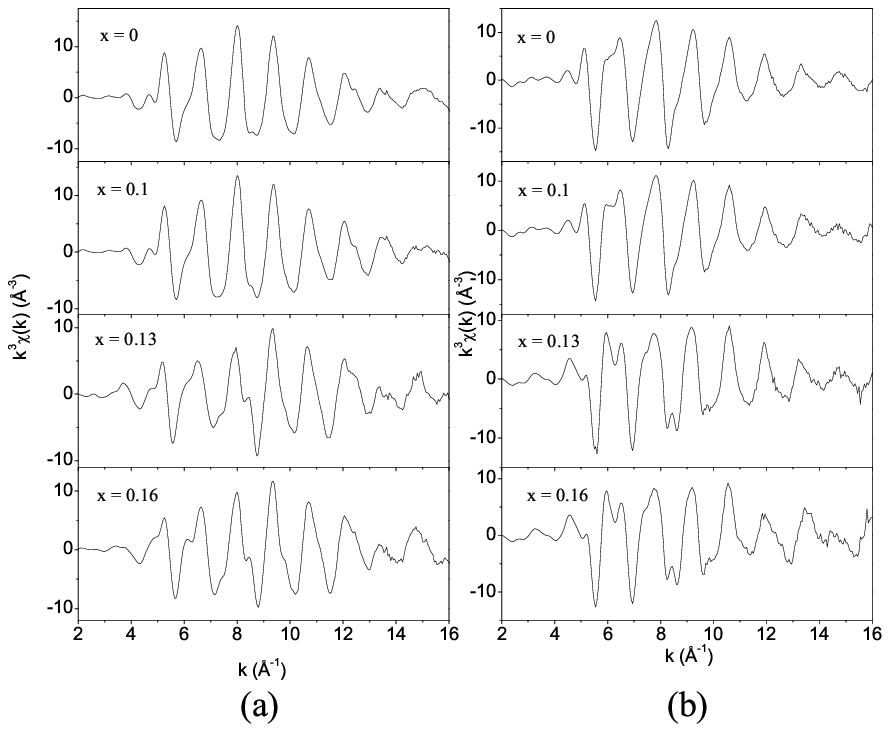, width=8cm, height=8cm}
\caption{\label{RT-all-k} The room temperature $k^3$ weighted $\chi (k)$ spectra of $x$ = 0, 0.1, 0.13, 0.16 samples for (a) Mn K edge (b) Ga K edge. These data were Fourier transformed in the range (2-15) (\AA$^{-1}$) for analysis.}
\end{figure}

\begin{figure}[h]
\epsfig{file=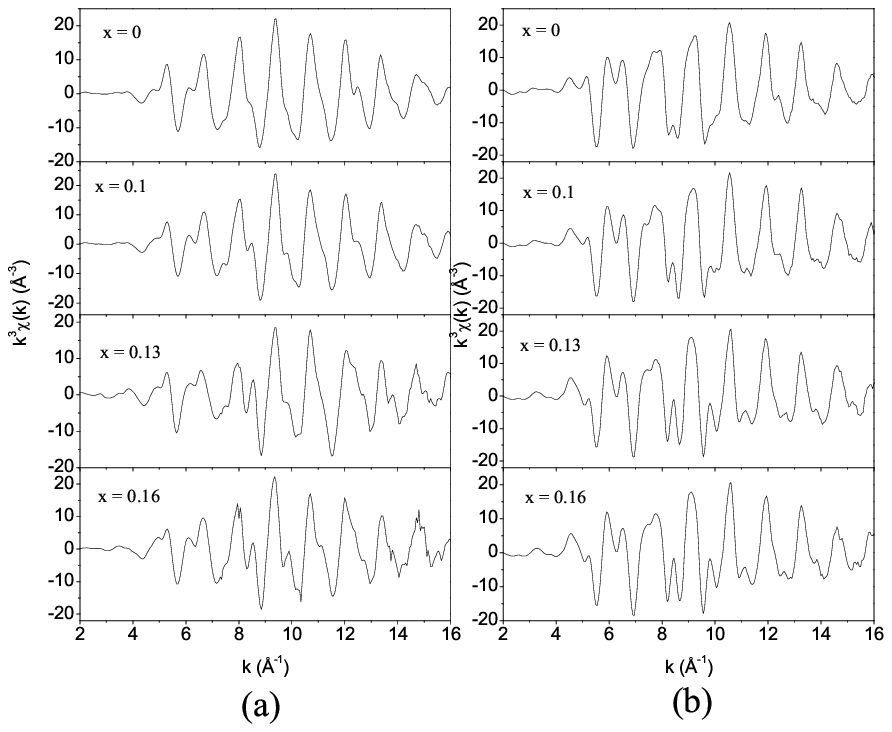, width=8cm, height=8cm}
\caption{\label{LT-all-k} The low temperature $k^3$ weighted $\chi(k)$ spectra of the indicated samples for (a) Mn K edge and (b) Ga K edge. The data in the range (2-15) (\AA$^{-1}$) was Fourier transformed for analysis.}
\end{figure}

\begin{figure}[h]
\epsfig{file=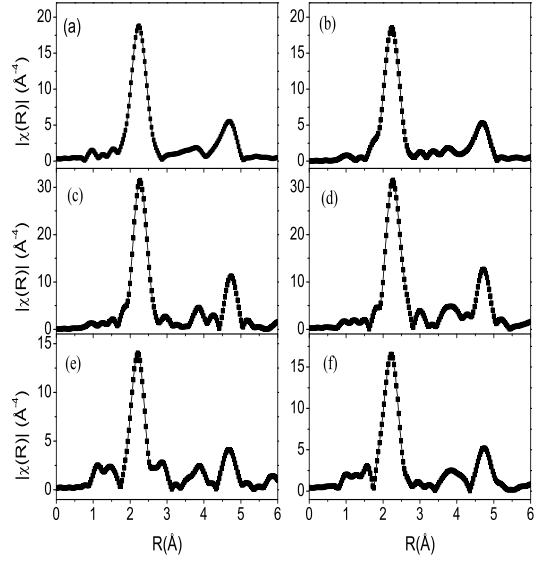, width=8cm, height=9cm}
\caption{\label{mag-R-all} The magnitude of Fourier Transform spectra of Mn K-edge EXAFS in the  austenitic phase (room temperature) for (a) $x$ = 0 (b) $x$ = 0.1 and in the martensitic phase for (c) $x$ = 0, (d) $x$ = 0.1, (e) $x$ = 0.13 and (f) $x$ = 0.16.}
\end{figure}

Absorbers for the EXAFS experiments were made by spreading very fine powder on a scotch tape avoiding any sort of sample inhomogeneity and pin holes. Small strips of the sample coated tape were cut and were held one on top of other. Enough number of such strips were adjusted to give absorption edge jump, $\Delta\mu x \le 1$. EXAFS at Mn and Ga K-edges were recorded in the transmission mode at the EXAFS-1 beamline at ELETTRA Synchrotron Source using Si(111) as monochromator. The measurements were carried out at room temperature and liquid Nitrogen temperature (henceforth called RT and LT respectively). The incident and transmitted photon energies were simultaneously recorded using gas-ionization chambers filled with mixtures of He-N$_2$ for Mn edge and Ar-N$_2$ for Ga edge. Measurements were carried out from 300eV below the edge energy to 1000eV above it with a 5eV step in the pre-edge region and 2.5eV step in the EXAFS region. At each edge, three scans were collected for each sample. Data analysis was carried out using IFEFFIT suite wherein theoretical fitting standards were computed with ATOMS and FEFF6 programs \cite{rav, zab} and fitting was done using FEFFIT program \cite{new}. In the present series although Mn content changes from 1 to 0.84, this change amounts to less than one atom per unit cell. Furthermore, the atomic number of Mn (Z =  25) and that of Ni (Z = 29) being similar, the x-ray scattering amplitudes and phase functions will not be drastically different and EXAFS would be insensitive to such a substitution. Therefore theoretical standards were calculated for stoichiometric Ni$_2$MnGa and were fitted to EXAFS data of all the samples irrespective of Mn content. The $k^3$-weighted $\chi(k)$ spectra at Mn and Ga K-edges in all the samples at RT and LT are shown in Fig.\ref{RT-all-k} and Fig.\ref{LT-all-k} respectively. These spectra reflect the good quality of data up to 15\AA$^{-1}$. The Fourier transform (FT) magnitude in $R$ space of the $k^3$ weighted Mn K-edge EXAFS at RT and LT are shown in Fig.\ref{mag-R-all}.

\section{\label{sec:level1}Results}
\subsection{\label{sec:level2}Austinitic Phase}
\begin{figure}[h]
\epsfig{file=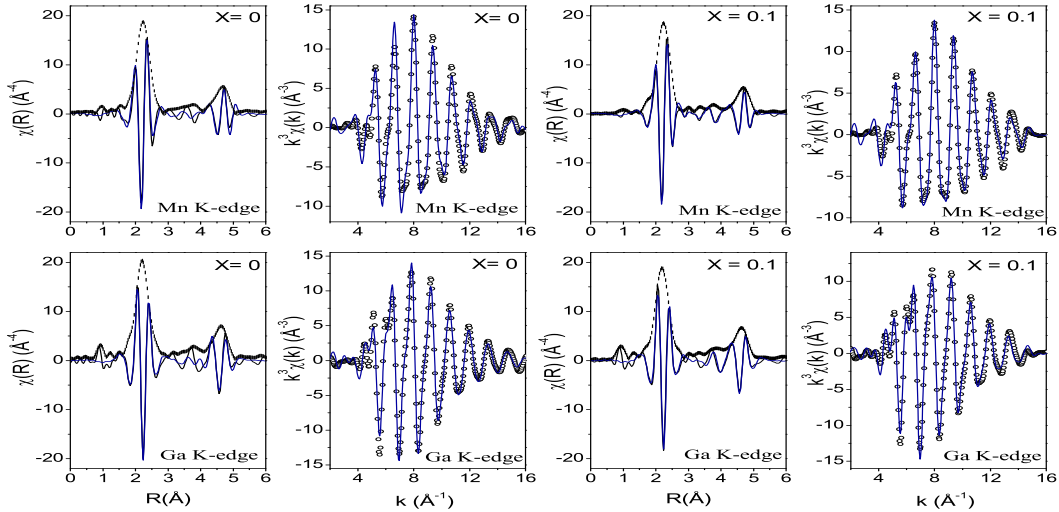, width=16cm, height=8cm}
\caption{\label{xafs-aus} Magnitude and real component of FT of EXAFS spectra in R space and real component of FT in the back transformed k space for Mn and Ga K-edge in $x$ = 0, 0.1 at room temperature. The fitting to the data are shown in blue line.}
\end{figure}

At room temperature the samples with $x$ = 0 and 0.1 are in the austinitic phase. Therefore, the EXAFS spectra of these samples recorded at Mn and Ga K-edges was fitted using common set of variable parameters with Fm3m space group and lattice constant 5.825\AA. In this model, the correction to the path lengths was refined with a constraint, 
$$ \delta R = \delta r_1 \times \frac{R_{eff}}{R_{nn1}}$$ 
where $R_{nn1}$ is the nearest neighbour distance, kept fixed to 2.5223\AA~ obtained from the lattice constant, $R_{eff}$ is the calculated bond length obtained from FEFF and $\delta r_1$ is the change in first neighbour distance. This approach reduces the number of variable parameters in the fit. The thermal mean-square variation in the bond lengths, $\sigma^2$ were varied independently for each path considered in the fit. The fitting was carried out in $R$-space in the range 1\AA~ to 5\AA~ using four single scattering (SS) paths and one linear multiple scattering (MS) path along the body diagonal of the initial cubic cell. The magnitude and real component of FT of $k^3 \chi (k)$ for Mn and Ga edge data are shown in Fig.\ref{xafs-aus}. As can be seen from the figure, the fits are quite satisfactory. The bond distances and the final fitted parameters obtained are presented in the Table \ref{tab-aus}. 

It is seen that there is a discrepancy in the Mn-Ga/ Ga-Mn bond distance for $x$ = 0.1 sample. The $\sigma^2$ values for this bond are also quite different. The reason for this anomaly could be the proximity of its martensitic transformation temperature (T$_M$ = 285K) to the temperature of EXAFS measurement (295K). It is well known in literature that the martensitic transformations are preceeded by a pre-transformation effects like softening of phonon modes and anomalies in elastic constants. Inelastic neutron scattering studies have evidenced such anomalies in Ni-Mn-Ga alloys \cite{Zheu, stur}. The T$_M$ for $x$ = 0.1 sample being only 10K below the room temperature, pre-transformational effects would be much intense here, giving rise to anamolies in $\sigma^2$ and causing discrepancies in bond distance of the near-neighbour atoms.

\subsection{\label{sec:level2}Martensitic Phase}
\subsubsection{\label{sec:level3}Samples with composition $x$ = 0, 0.1}
Fig.\ref{mag-R-all}(c) and (d) demonstrates the low temperature EXAFS in $R$ space at Mn edge for $x$ = 0 and 0.10 samples present in their martensitic phase. In the range R = 2.5 - 5.0\AA~ a difference in spectral signatures of the two alloys in the LT and RT data is quite evident. This can be attributed to the lowering of symmetry from the parent cubic structure upon undergoing the martensitic transition. Consequently, the EXAFS analysis was carried out using a tetragonal structure with c/a $<$ 1 \cite{mart}. The $\sigma^2$ values obtained from the austinitic phase served as starting parameters and $\delta R$ parameters were varied independently. The co-ordination number for each path was kept fixed to its crystallographic value. As per the model, for Mn as the absorbing atom, the first peak in the range R = 1.5 to 3.2\AA~ is due to the contribution from 8 Ni atoms at 2.52\AA,~ 2 Ga atoms at 2.78\AA~ and 4 Ga atoms at 2.96\AA. However, the $\sigma^2$s for Mn-Ga paths obtained from this fitting differs vastly from each other with values of 0.003\AA$^2$ and 0.03\AA$^2$ respectively. Generally, at such close bond lengths, a large variation in $\sigma^2$ especially for the bonds involving same type of atoms is not expected. Therefore it indicates that there is a large spread in the bond distance of longer Ga neighbour and/or a different distribution of Ga atoms around Mn in the second and third shells.  Thus the model was supplemented by carrying out fits with different combinations of 6 Ga neighbours distributed in the two shells. In each of these fits, $\sigma^2$ was varied keeping the coordination number fixed to a particular distribution. Best fit was obtained for 4 and 2 Ga atoms in the second and third shell respectively and the $\sigma^2$ obtained are presented in table \ref{tab-mart}. Thereafter, such re-grouping of bond lengths had to be incorporated for all the subsequent SS paths. Here, the important aspect brought out by the analysis is the change in atomic coordinations leading to distribution of the bond lengths. This observation reflects the different arrangements of atoms in different crystal planes in the martensitic phase. In other words, the constituent atoms have been displaced over varied distances giving rise to modulations in the crystal planes in trying to maintain volume conservation - an essential criteria for martensitic transition. Low temperature neutron diffraction studies on Ni$_2$MnGa by Brown {\it et.al} \cite{brow14} have reported a modulated structure for Ni$_2$MnGa from which a similar grouping of bond length distribution of Ga atoms around Mn can is obtained.

\begin{figure}[h]
\epsfig{file=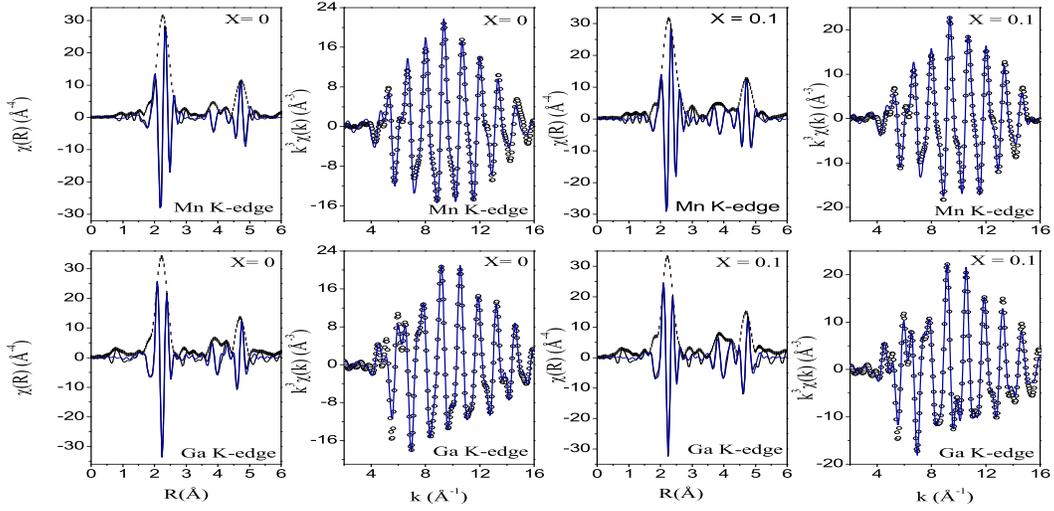, width=16cm, height=8cm}
\caption{\label{LT-200-R} The low temperature Mn and Ga K-edge FT magnitude and real component of EXAFS spectra in R space and real component of FT in the back transformed k space $x$ = 0, 0.1. The best fit to the data are shown as blue line.}
\end{figure}

Refinement of Ga K-edge data also presents a similar situation. The parameters extracted from both the edges are presented in Table \ref{tab-mart} and FT fittings in $R$ space are shown in the Fig.\ref{LT-200-R}. The bond distribution with 4 and 2 coordination of Mn atoms are present around the central Ga atom at an averarge distance of 2.791\AA~ and 3.065\AA~ respectively. It is seen from the table that the third bond distance with Ga as central atom is larger by about 0.1\AA~ as compared to that with Mn as the central atom. Also the $\sigma^2$ values obtained from Ga EXAFS for Ga-Mn bonds are higher than those obtained from Mn EXAFS (refer table \ref{tab-mart}). The physical significance of these observations is that the Ga atoms have a smaller amplitude of displacement from its crystallographic position in comparison to Mn. In other words, Ga atoms are sluggish and do not get much displaced in undergoing a martensitic transition leading to a stronger hybridization between Ga-Ni in the martensitic phase.

Another important observation here is the discrepancy in the bond distance of the of the MS path described in Table \ref{tab-mart}. Being a linear path along the body diagonal of the initial cubic cell, the length of this path should be the sum of Mn-Ni and Ga-Ni bond lengths. Apparently, this condition is not satisfied in the LT data which implies that the MS path is no longer linear. It is this signature that once again brings in the prominence of movement of atoms in the martensitic phase. The distortions in atomic positions further results in ``dimpling'' of crystal planes as evidenced by non-linearity of the MS path giving rise to modulated structures. In short, it is seen from our RT and LT EXAFS analysis of $x$ = 0, 0.10 samples that the martensitic transformation causes the atoms to displace from its initial positions by varied amplitudes with least displacement of Ga atoms causing local distortions. These local distortions gives rise to modulations and may eventually lead to long-range ordering of unit cells over many atomic planes. 

\subsubsection{\label{sec:level3}Samples with composition $x$ = 0.13, 0.16}
\begin{figure}[h]
\epsfig{file=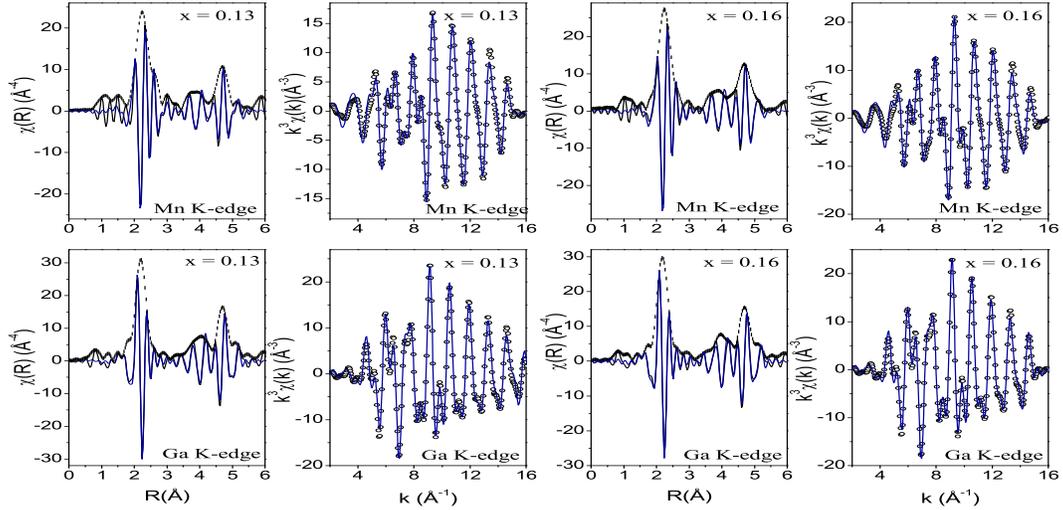, width=16cm, height=8cm}
\caption{\label{LT-213-R} Low temperature, $k^3$ weighted, FT EXAFS spectra in R space and real component of FT in the back transformed k space for Mn and Ga K-edge for $x$ = 0.13 and 0.16 samples. The blue line indicates best fit to the data.}
\end{figure}

The samples with $x$ = 0.13, 0.16 are martensitic at room temperature. As can be seen from Fig.\ref{mag-R-all}(e) and (f), the RT spectral signatures of these samples show subtle differences in comparison to the LT spectra for $x$ = 0, 0.1. Thus a tetragonal structural model with $c/a >$ 1 was employed for interpretation of the spectra \cite{wedel}. In these samples, the RT and LT spectra are quite similar and hence the FT fittings in $R$ space are presented for the low temperature data alone in Fig.\ref{LT-213-R}. The bond distances obtained are presented in Table \ref{tab-mart}. It is seen that the bond distances obtained from the two edges show notable difference, especially in the first shell. If one considers the difference between Mn-Ni and Ga-Ni bond distances alone, there is a change of about $\sim$ 0.016\AA. Both the central atoms, Mn and Ga can be viewed to be at the body centered position of a reduced tetragonal structure formed by 8 Ni atoms. A non-uniformity in their bond distance with Ni of the order of 10$^{-2}$\AA ~ is unexpected and hints toward the microscopic changes influencing the formation of the macroscopic modulated phases. Also, the second and third shell distances obtained from the two edges are significantly different. 

Furthermore, as can be seen from table\ref{tab-mart}, Ga-Mn($\sim$ 3\AA)~ bond distance is shorter that the Mn-Ga($\sim$ 3.2\AA)~ bond distance. This is exactly opposite to the trend observed in LT spectra of $x$ = 0, 0.1 where Ga-Mn = 3.06\AA~ and Mn-Ga = 2.96\AA. However, the $\sigma^2$ of Ga-Mn bond is much larger (0.03 \AA$^2 >$  0.01 \AA$^2$) than that of Mn-Ga bond. This typical behaviour of higher $\sigma^2$ for Ga-Mn bonds is prevalent in all the samples in martensitic phase. This result is critical because it is a direct indication of movement of constituent atoms from their crystallographic positions with Ga having the least amplitude of displacement. Thus, when viewed from Ga K-edge, the local structure seems to be much distorted with higher amplitudes of displacements for other constituent atoms and therefore a higher value of $\sigma^2$ for the respective bond. 
 
\section{\label{sec:level1}Discussion}
The local structural study of Ni$_{2+x}$Mn$_{1-x}$Ga alloys in the austinitic and martensitic phases by EXAFS at Mn and Ga K-edge enable us to identify the microscopic changes influencing the formation of the macroscopic modulated structure. Based on our analysis the following results were obtained:
\begin{itemize}
\item Higher values of thermal mean square vibration, $\sigma^2$, for Ga K-edge analysis in comparison to Mn K-edge in the martensitic phase for $x$ = 0, 0.10 samples. 
\item Different values of bond distances for the same pair of atoms (Ga-Mn) in the martensitic phase of all samples. 
\item A difference of 0.016\AA~ between Mn-Ni and Ga-Ni bond distance in $x$ = 0.13, 0.16.
\end{itemize}

Firstly, the parameters obtained for the room temperature EXAFS data for $x$ = 0, 0.1 are in line with those expected for the austenitic structure. The low temperature EXAFS spectra, for these samples have features different to those of the room temperature and carry information about the martensitic phase. The displacement of atoms occurring upon the structural change is reflected through higher values of $\sigma^2$. This argument is further supported by change in coordination numbers to 4+2 and re-grouping of the two Mn-Ga/Ga-Mn bonds. Thus the local symmetry breaking upon structural phase transition leads to the movement of constituent atoms. Such a modulated structure of Ni$_2$MnGa has been determined experimentally\cite{brow14} and effect of shuffling of atoms on the physics of martensitic transformation has also been studied theoretically\cite{zay-prb-68, zay-prb-72}. On inspection of the parameters extracted from our analysis, it is seen that the $\sigma^2$ values are higher in case of Ga EXAFS than Mn EXAFS for the same Mn-Ga bond in the martensitic phase. This clearly indicates a spread or distribution of Mn-Ga bonds wherein Mn atoms moves more freely in comparison to Ga. This observation is further substantiated by a higher average bond length obtained with respect to Ga as central atom as compared to that obtained from Mn. Therefore there is a larger spread in Ga-Mn distance than Mn-Ga. In other words, Ga atoms are sluggish and have a smaller amplitude of displacement than other constituent atoms forming the alloy.

\begin{figure}[h]
\epsfig{file=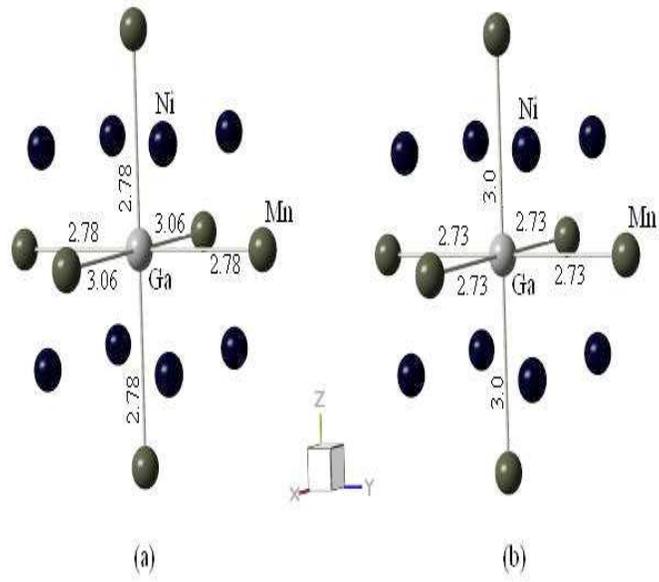, width=10cm, height=8cm}
\caption{\label{loc-str} The local enviorment around the central Ga atom in martensitic phase (a) for $x$ = 0.1 and (b) for $x$ = 0.13.}
\end{figure}

Thermal and stress induced martensitic transitions in Ni-Mn-Ga single crystals has been studied in the past \cite{mart} wherein, it was shown that the sample in martensitic phase undergoes another stress induced transition from a structure with $c/a < 1$ to that with $c/a > 1$.   A similar transition is seen here in Ni$_{2+x}$Mn$_{1-x}$Ga when $x$ changes from 0.1 to 0.13. A simplistic view, as obtained from our EXAFS analysis, of local structure from Ga as central atom in the martensitic phase is as shown in Fig. \ref{loc-str}. In the case of $x = 0.1$ wherein $c/a$ is $<$ 1, the effect of  modulations are shown to be about the $b$ axis. The shorter Mn-Ga bond lengths are therefore depicted along the $b$ and $c$ axes while the longer one is shown along the $a$ axis. For $x = 0.13$, as described above, EXAFS can only be fitted to a structure with $c/a > 1$. Therefore in this case the longer Mn-Ga bond length is shown along the $c$ axis and the shorter ones along $a$ and $b$ axes. A comparison of magnitudes of bond lengths clearly shows that in $x = 0.13$ a rotation about $b$ axis transforms the structure similar to that of $x = 0.1$. Therefore the two martensitic structures are similar under rotation of cartesian axes. 
  
In order to further elucidate the local structure in Ni-Mn-Ga system, we look at the nearest neighbour interaction of the absorbing atoms. As mentioned in the previous section, there is a notable difference in the Mn-Ni and Ga-Ni bond lengths determined from the low temperature data for $x$ = 0.13, 0.16. Comparing the estimates obtained, it is seen that Mn-Ni bond distance is higher than the Ga-Ni distance. The difference is greater than experimental resolution and demonstrates the locally distorted environment around Mn and Ga atoms leading to an electronic structure that is different in the austinitic and martensitic phases. Ni atoms reside inside the interlocking tetrahedral cages formed by  Mn and Ga atoms. A shorter Ga-Ni bond means that Ga is more strongly bound to Ni than Mn. Thus the Ga tetrahedra is distorted in a way that allows more space for the movement of Mn atoms. Therefore Mn atoms have higher amplitude of displacement from its crystallographic position than Ga. These tetrahedral distortions lead to re-distribution of electrons and is perhaps the root cause of band Jahn-Teller transitions observed in such alloys\cite{fuji, brow11}. Such a Ni-Ga hybridization has also been anticipated theoretically\cite{zay-prb-72}. These calculations yeild energetically favorable hybrid states formed by Ga and Ni giving rise to a peak in the spin-down electronic density of states at the Fermi level. 

\section{\label{sec:level1}Conclusions}
In this work, we have carried out a comparative analysis of changing local structures in Ni$_{2+x}$Mn$_{1-x}$Ga alloys upon undergoing martensitic transition. EXAFS measurements at Mn K-edge and Ga K-edge at room temperature and liquid Nitrogen temperature were carried out. The most significant feature of our analysis is the difference in the Mn-Ni and Ga-Ni bond length. A shorter Ga-Ni bond implies an increased Ga-Ni hybridization in comparision to Mn-Ni in the martensitic phase. The present study is a direct experimental evidence for such hybridization which is seen more clearly in $x$ = 0.13 and 0.16 samples. The differences in the Mn-Ni and Ga-Ni bond lengths leads to the distortion of the two tetrahedra formed by Mn-Ni and Ga-Ni. The increased hybridization would probably lead to re-distribution of electrons causing a band Jahn-Teller effect. 

It is also seen that the constituent atoms of Ni-Mn-Ga system displace from their crystallographic positions by varying amplitudes in the martensitic phase. The Ga atoms seems to displace over very small amplitudes. The uneven movement of the constituent atoms gives rise to dimpling of atomic planes and may eventually lead to modulated structures.

\acknowledgements
Authors gratefully acknowledge financial assistance from Department of Science a nd Technology, New Delhi, India and ICTP-Elettra, Trieste, Italy for the proposal 2004646 and 2005743. Thanks are also due to Prof. G. Vlaic and Dr. Luca Olivi for help in EXAFS measurements and useful discussions. P.A.B. would like to thank Council for Scientific and Industrial Research, New Delhi for Senior Research Fellowship.

\begin{table}[h]
\caption{Results of the fits to the room temperature Mn and Ga edge data of $x$ = 0, 0.10. R refers to the bond length and $\sigma^2$ is the thermal mean square variation in the bond length. The fittings were carried out in $k$ range: (2- 15)\AA$^{-1}$ with $k$-weight: 3 and $R$ range: (1-5)\AA. Figures in parantheses indicate uncertainity in the last digit.}
\vspace{0.2cm}
\centering
Mn K-edge
\begin{ruledtabular}
\begin{tabular}{rcccc}
Atom and & \multicolumn{2}{c}{$x$ = 0} & \multicolumn{2}{c}{$x$ = 0.1}\\
\cline{2-5}
Coord. No. & R (\AA) & $\sigma^2$ (\AA$^2$) & R (\AA) & $\sigma^2$ (\AA$^2$) \\
\hline 
Ni1 $\times$ 8 & 2.519(8) & 0.0081(3) & 2.518(2) & 0.0080(2)\\
Ga1 $\times$ 6 & 2.909(3) & 0.03(1) & 2.795(8) & 0.014(1)\\
Mn1 $\times$ 12 & 4.114(4) & 0.029(9) & 4.20(2) & 0.021(3)\\
Ni2 $\times$ 24 & 4.824(5) & 0.019(3) & 4.85(1) & 0.019(2)\\
Ga2 $\times$ 16 & 5.038(5) & 0.007(1) & 4.900(8) & 0.008(1)\\
MS\footnote{Mn$\rightarrow$Ga3$\rightarrow$Ni1$\rightarrow$Mn} $\times$ 8 & 5.038(5) & 0.0097(6) & 5.02(1) & 0.018(2)\\
\hline
&&&&\\
\multicolumn{5}{c}{Ga K-edge}\\
\hline
\hline
Atom and & \multicolumn{2}{c}{$x$ = 0} & \multicolumn{2}{c}{$x$ = 0.1}\\
\cline{2-5}
Coord. No. & R (\AA) & $\sigma^2$ (\AA$^2$) & R (\AA) & $\sigma^2$ (\AA$^2$) \\
\hline
Ni1 $\times$ 8 & 2.512(2) & 0.0077(2) & 2.516(3) & 0.0082(4)\\
Mn1 $\times$ 6 & 2.901(2) & 0.030(6) & 2.88(5) & 0.026(7)\\
Ga1 $\times$ 12 & 4.103(3) & 0.022(4) & 4.23(4) & 0.017(5)\\
Ni2 $\times$ 24 & 4.811(4) & 0.015(1) & 4.79(1) & 0.016(2)\\
Mn2 $\times$ 16 & 5.025(3) & 0.017(7) & 4.90(1) & 0.007(1)\\
MS\footnote{Ga$\rightarrow$Mn3$\rightarrow$Ni1$\rightarrow$Ga} $\times$ 8 & 5.025(3) & 0.014(1) & 5.15(3) & 0.023(4)\\
\end{tabular}
\end{ruledtabular}
\label{tab-aus}
\end{table}

\begin{table*}[h]
\caption{Results of the fits to the low temperature Mn and Ga edge data of $x$ = 0, 0.10, 0.13 and 0.16. R refers to the bond length and $\sigma^2$ is the thermal mean square variation in the bond length. The fittings were carried out in $k$ range: (2- 15)\AA$^{-1}$ with $k$-weight: 3 and $R$ range: (1-5)\AA. Figures in parantheses indicate uncertainity in the last digit.}
\vspace{0.2cm}
\centering
Mn K-edge
\begin{ruledtabular}
  \begin{tabular}{lllllllll}
Atom and & \multicolumn{2}{c}{$x$ = 0} & \multicolumn{2}{c}{$x$ = 0.1} & \multicolumn{2}{c}{$x$ = 0.13} & \multicolumn{2}{c}{$x$ = 0.16}\\
 \cline{2-9}
Coord. No. & R (\AA) & $\sigma^2$ (\AA$^2$) & R (\AA) & $\sigma^2$ (\AA$^2$) & R (\AA) & $\sigma^2$ (\AA$^2$) & R (\AA) & $\sigma^2$ (\AA$^2$) \\
\hline
Ni1 $\times$ 8 & 2.518(3) & 0.0057(4) & 2.518(5) & 0.0056(5) & 2.528(2) & 0.0060(2) & 2.523(2) & 0.0051(1)\\  
Ga1 $\times$ 4 & 2.780(6) & 0.0043(5) & 2.768(8) & 0.0037(7) & 2.740(3) & 0.0045(3) & 2.739(5)& 0.0054(4)\\
Ga2 $\times$ 2 & 2.96(3) & 0.008(3) & 2.95(2) & 0.009(2) & 3.12(2) & 0.010(3) & 3.23(4) & 0.012(5)\\
Mn1 $\times$ 4 & 3.96(3) & 0.009(3) & 3.93(3) & 0.009(3) & 3.907(3) & 0.012(4) & 3.89(2) & 0.008(2)\\
Mn2 $\times$ 8 & 4.19(1) & 0.009(2) & 4.19(1) & 0.009(2) & 4.22(1) & 0.009(1) & 4.23(2) & 0.011(2)\\
Ni2 $\times$ 16 & 4.69(1) & 0.011(1)& 4.66(1) & 0.009(1) & 4.61(1) & 0.013(2) & 4.613(9) & 0.009(1)\\
Ni3 $\times$ 8 & 4.90(1) & 0.006(1) & 4.905(8) & 0.0040(7) & 5.364(8) & 0.0044(8) & 5.327(8) & 0.0044(8)\\
MS\footnote{Mn$\rightarrow$Ga3$\rightarrow$Ni1$\rightarrow$Mn} $\times$ 16 & 5.068(6)& 0.0095(7)& 5.056(8) & 0.0097(8) & 5.088(4) & 0.0078(4) & 5.075(3) & 0.0068(3)\\
\hline
&&&&&&&&\\
\multicolumn{9}{c}{Ga K-edge}\\
\hline
\hline
Atom and & \multicolumn{2}{c}{$x$ = 0} & \multicolumn{2}{c}{$x$ = 0.1} & \multicolumn{2}{c}{$x$ = 0.13} & \multicolumn{2}{c}{$x$ = 0.16}\\
 \cline{2-9}
Coord. No. & R (\AA) & $\sigma^2$ (\AA$^2$) & R (\AA) & $\sigma^2$ (\AA$^2$) & R (\AA) & $\sigma^2$ (\AA$^2$) & R (\AA) & $\sigma^2$ (\AA$^2$) \\
\hline
Ni1 $\times$ 8 & 2.5111(8) & 0.00431(8) & 2.512(2) & 0.0044(2) & 2.512(1) & 0.0039(1) & 2.511(1) & 0.0042(1)\\
Mn1 $\times$ 4 & 2.791(4) & 0.0078(5) & 2.776(8) & 0.0062(5) & 2.725(4) & 0.0063(5) & 2.722(5) & 0.0067(6)\\
Mn2 $\times$ 2 & 3.065(2) & 0.012(2) & 3.06(2) & 0.013(4) & 3.0(2) & 0.04(3) & 3.0(1) & 0.03(2) \\
Ga1 $\times$ 4 & 3.97(1) & 0.009(2) & 3.93(2) & 0.008(2) & 3.87(2) & 0.008(2) & 3.85(2) & 0.009(2)\\
Ga2 $\times$ 8 & 4.215(8) & 0.0076(8) & 4.214(9) & 0.0079(9) & 4.25(1) & 0.009(1) & 4.248(7) & 0.009(1)\\
Ni2 $\times$ 16 & 4.706(8) & 0.0048(6) & 4.676(6) & 0.0081(7) & 4.614(7) & 0.0078(6) & 4.619(7) & 0.0083(7)\\
Ni3 $\times$ 8 & 4.889(3) & 0.0047(3) & 4.872(5) & 0.0033(4) & 5.313(8) & 0.0026(5) & 5.319(8) & 0.0025(5)\\
MS\footnote{Ga$\rightarrow$Mn3$\rightarrow$Ni1$\rightarrow$Ga} $\times$ 16 & 5.069(7) & 0.0111(9) & 5.111(5) & 0.0089(6) & 5.102(4) & 0.0051(3) & 5.106(2) & 0.0047(2)\\
\end{tabular}
\end{ruledtabular}
    \label{tab-mart}
\end{table*}

\end{document}